\documentclass[conference]{IEEEtran}
\usepackage{amsmath,amssymb}

\usepackage{amssymb}
\usepackage{bm}
\usepackage[dvipdfmx]{graphicx}
\usepackage{color}

\graphicspath{{figure/}}

\newcommand{\defeq}{\stackrel{\triangle}{=}}
\newtheorem{example}{Example}
\newtheorem{definition}{Definition}

\newtheorem{lemma}{Lemma}

\newtheorem{theorem}{Theorem}

\def\qed{\hfill $\Box$} 
\usepackage{setspace}

\begin{document}
\title{Subgraph Domatic Problem and Writing Capacity of Memory Devises with Restricted State Transitions} 
\author{
  \IEEEauthorblockN{Tadashi Wadayama$^*$, Taizuke Izumi$^*$, and Hirotaka Ono$^\dagger$}
  \IEEEauthorblockA{$^*$Nagoya Institute of Technology,   Japan\\
  $^\dagger$Kyusyu University,   Japan \\
      Email: wadayama@nitech.ac.jp, t-izumi@nitech.ac.jp, hirotaka@en.kyushu-u.ac.jp } 
}

\maketitle
\begin{abstract}
A code design problem
for memory devises with restricted state transitions is formulated as a 
combinatorial optimization problem 
that is called a {\em subgraph domatic partition} (subDP) problem.
If any neighbor set of a given state transition graph contains all the colors,
then the coloring is said to be valid.
The goal of a subDP problem is to find a valid coloring with the largest number of colors for
a subgraph of a given directed graph.  The number of colors in an optimal valid 
coloring gives the writing capacity of a given state transition graph.
The subDP problems are  computationally hard; it is proved to be NP-complete in this paper. 
One of our main contributions in this paper is to show the asymptotic behavior of 
the writing capacity $C(G)$ for sequences of dense bidirectional graphs,  that is given by 
$C(G)=\Omega(n/\ln n)$ where $n$ is the number of nodes.
A probabilistic method called  {\em Lov\'asz local lemma} (LLL) plays an essential  role 
to derive the asymptotic expression. 
\end{abstract}

\section{Introduction}

Recent advanced memory devises such as flash memory and phase change memory (PCM)
require appropriate coding for improving its write efficiency.
For example, coding schemes for flash memories \cite{Jiang}\cite{Bruck} \cite{Average} \cite{ILIFC}
can improve write efficiency so that lifetime of the flash memory is lengthened.
Several constrained coding schemes suitable for PCM 
has been presented \cite{pcm} \cite{Qin}. Rank modulation developed by Jiang et al. \cite{rankmodulation} that encodes 
a message onto a permutation of $n$-elements produced a number of technically and 
mathematically interesting problems. In the paper by En Gad et al. \cite{Eyal},
they considered a combinatorial problem to find a maximum size of 
the decomposition of a state space into several dominating sets of state transition graphs 
in order to determine the maximal rate of compressed rank modulation.

A problem to partition nodes of a given undirected graph into several 
non-overlapping dominating sets is called a {\em domatic partition problem} 
\cite{olddomatic}\cite{Feige}.
The domatic partition problem has been long known in the field of 
theoretical computer science.  The problem can be considered as a special 
type of a node coloring problem for an undirected graph. Solving the 
problem in an exact manner is considered to be computationally intractable
since it was proved that the domatic partition problem is an NP-hard problem \cite{Garey}.
Several efficient approximation algorithms has been developed and 
analyzed \cite{Feige} for the domatic partition problems.

Our aim of this paper is to provide  theoretical foundation 
to the problems called {\em subgraph domatic partition} (SubDP) problems
from a graph theoretic view point. 
In fact, the motivation of this paper came directly from the paper of En Gad et al. \cite{Eyal}.
In our problem setting, a directed graph $G$ that represents allowable state transitions 
of a memory devise is initially given. We also assume that a fixed size message should be 
written in the memory for each state transition. In order to construct an appropriate 
encoding and decoding functions with high coding rate,  we must solve 
an SubDP problem on $G$. A SubDP problem is a problem 
to find both of a subgraph of $G$ and its domatic partition with largest number of 
groups; {\it i.e.}, domatic coloring.
Although the SubDP problem is so closely related to the standard domatic partition problem, 
it has not been discussed in literatures as far as the authors know.

Our approach to the SubDP problem is summarized as follows.
We will  firstly give a rigid definition of the {\em writing capacity} for a directed 
graph that corresponds to the solution of an SubDP problem. This can be 
considered as an abstraction that transforms coding problems  to 
combinatorial problems on directed graphs.
We then invoke {\em Lov\'asz local lemma} (LLL) to derive 
a non-trivial lower bound on the writing capacity. LLL is a well-known probabilistic method \cite{alon}
that is often used to show an existence of certain configuration satisfying all the requirements.
The lower bound based on LLL reveals asymptotic behaviors of 
the writing capacities $C(G)$ of bidirectional graphs that follows $C(G) = \Omega(n/\ln n)$ 
for a sequence of dense graphs where $n$ is the number of nodes.
Furthermore, we will show that $C(G) = \Theta(n)$ for a sequence of extremely dense graphs
by using {\em Tur\'an's lemma} that is a fundamental lemma in extremal graph theory.

\section{Preliminaries}

In this section, notation used throughout the paper 
and definition related to the SubDP problems
will be given.

\subsection{Notation}

Let $G = (V,E)$ be a directed graph. For a node $v \in V$,
the outbound degree and inbound degree of $v$ are
denoted by $d_{out}(v)$ and $d_{in}(v)$, respectively.
The average outbound degree $\epsilon_{out}(G)$ of 
$G$ is given by
\begin{equation}
\epsilon_{out}(G)= \frac{1}{|V|}\sum_{v \in V} d_{out}(v).
\end{equation}
Note that $\epsilon_{out}(G) = {|E|}/{|V|}$ holds due to the relation 
$\sum_{v \in V} d_{out}(v) = |E|$. The symbols $\delta_{out}$ and $\Delta_{out}$
represent minimum and maximum outbound degree of $G$, respectively.
In a similar manner, the symbols $\delta_{in}$ and $\Delta_{in}$
are the minimum and maximum inbound degree.
If a directed graph $G=(V,E)$ satisfies 
$
(u,v) \in E \leftrightarrow (v,u) \in E,
$
then the graph $G$ is said to be a {\em bidirectional graph}.
In this paper, we use the term ``bidirectional'' instead of ``undirected'' 
in order to show that there are two directed edges between $u$ and $v$ explicitly.

\subsection{Basic assumptions}

Assume that a directed graph $G = (V,E)$, 
called a {\em state transition graph}, is given. 
The memory devise $D$ associated to the graph $G$ can store
any $v \in V$ as its state. The state of $D$ can be observed from outside.
We can change the state of $D$ from $s \in V$ to $s' \in V$ if
the directed edge $(s,s')$ is included in $E$.
Let ${\cal M} \defeq \{1,2,\ldots, M \}$ called an {\em message alphabet}.
We assume the following scenario.
When a request to write a message $m \in {\cal M}$ comes,  we are allowed to 
change the state of $D$ only once along with 
an edge in $E$. After the state transition for writing $m$, 
it is required that a written message in $D$ must be 
correctly recovered by reading the state of $D$.
An appropriate pair of an encoding and a decoding function satisfying these assumptions
is required to realize a memory system based on the devise $D$.

\subsection{Encoding function}

The main task of our encoder is to output a next state $s'$ of $D$ from 
an input pair of a message $m$ and a current state $s$. The important constraint 
imposed on the encoder is $(s,s')$ is in $E$ or $s'=s$. This means that only 
a state transition (or no transition) consistent with the state transition graph $G$
is allowed. The following formal definition of encoding functions certainly includes
this consistency condition. 

Let $\tilde G = (\tilde V,  \tilde E)$ be a non-empty subgraph of $G$.
A function $\phi: \tilde V \times {\cal M} \rightarrow \tilde V $ satisfying 
\begin{equation} \label{encoding_cond}
\forall s \in \tilde V, \forall m \in {\cal M}, \quad  (s, \phi(s,m)) \in \tilde E \mbox{ or } s = \phi(s,m)
\end{equation}
is called an {\em encoding function}. 
It should be remarked that an encoding function does not need to use all the states in $V$.
An appropriate choice of $\tilde V \subseteq V$ may increase the writing capacity of $D$
defined later. 

An encoding process is described as follows. An encoder receives a message 
$m  \in {\cal M}$ to be stored in $D$ as an input. The first task of the encoder 
is to fetch the current state of $D$. Let $s  \in \tilde V$ be the current state stored in $D$.
We then apply the encoding function to the current state $s$ and the input message $m$
and obtain the next state $s' = \phi(s,m)$. The last task of the encoder is 
to change the state of $D$ from $s$ to $s'$.

Assume that $m_1, m_2, m_3, \ldots \in {\cal M}$ are a message sequence that 
is to be encoded sequentially and that $s_0$ is the initial state of $D$.
In this case, we obtain the state sequence 
$
s_0, s_1 = \phi(s_0, m_1), s_2 = \phi(s_1, m_2),  \ldots
$
that can be considered as a walk trajectory on the state transition graph $G$.
From this encoding process and the definition of the encoding function, 
it is evident that the state $s_i$ 
is contained in $\tilde V$ for any  $i (i \ge 1)$ if the initial state of $D$ is in $\tilde V$.
In the following discussion, we assume that $s_0$ is a state in $\tilde V$.

\subsection{Decoding function}

A current written message in $D$ must be correctly recovered 
when we want to read the stored message. This condition naturally leads to 
the following definition of the decoding function corresponding to an encoding function.

Let $\phi$ be an encoding function associated with the subgraph 
$\tilde G = (\tilde V, \tilde E)  \subseteq G$.
A function $\psi: \tilde V \rightarrow {\cal M}$ satisfying 
\begin{equation} \label{decodingcond}
\forall s \in \tilde V, \forall m \in {\cal M}, \quad  \psi(\phi(s,m)) = m
\end{equation}
is called an {\em decoding function}. 

A decoding process is simple. A decoder first fetch the state of $D$ 
when a read request comes. Then, the decoder applies the decoding function 
to the retrieved state $s$ and obtain a recovered message $m = \psi(s)$.
The condition (\ref{decodingcond}) guarantees that any stored message can be
correctly recovered.

\subsection{Writing capacity}

It is intuitively evident that there is a close relationship between 
topology of a given state transition graph and allowable size of 
a message alphabet.  In the following definition, we will introduce a measure 
that quantifies how much information can be written in a single state transition.
\begin{definition}[Writing capacity]
Let $G$ be a directed graph.
If there exist $\tilde G \subseteq G$ and a corresponding encoder-decoder pair $(\phi, \psi)$ satisfying both of the encoding condition (\ref{encoding_cond}) and the decoding 
condition (\ref{decodingcond}) for a given pair $(G, {\cal M})$, 
then the pair $(G, {\cal M})$ is said to be achievable.
The writing capacity $C(G)$ is defined by
\begin{equation}
C(G) \defeq \max \{|{\cal M}| :  (G,{\cal M}) \mbox{ is achievable} \}.
\end{equation}
\end{definition}

Although the definition of the writing capacity is fairly simple,  the evaluation of 
this quantity for a given graph $G$ is not trivial in general. 
It might be natural to consider the following problems on the writing capacity;
(1) how to evaluate the writing capacity and how to design encoder and decoder systematically, 
(2) worst case computational complexity to evaluate $C(G)$, (3) general upper and lower bounds, 
(4) finding relationships between the writing capacity and graph parameters (e.g., 
maximum degree, degree profile, etc.). In the following sections in this paper, some of these 
problems are to be discussed.

\subsection{Subgraph domatic partition}

In the following sections, we will give several lower and upper bounds on the
writing capacity. Before going into the discussions on such bounds, 
it might be beneficial to obtain the intuition on the problem by observing 
the following interpretation to a coloring problem. 

Assume that a subgraph $\tilde G \subseteq G$ is given.
An $\ell$-coloring is a map from $\tilde V$ to $[1, \ell]$, where the 
notation $[a,b]$ represents the set of consecutive integers from $a$ to $b$.
Let $c: \tilde V \rightarrow [1,\ell]$ be an $\ell$-coloring of $\tilde G$. 
For $v \in \tilde V$, $c(v) \in [1, \ell]$ represents its color.

The neighbor set of $v \in \tilde V$ is defined by
\begin{equation}
N(s)  \defeq \{s \} \cup  \{s' \in \tilde V \mid (s,s') \in \tilde E \}.
\end{equation}
Consistency between a decoding function $\psi$ and an encoding function $\phi$ 
can be translated into the following condition.
If an $\ell$-coloring $c$ on $\tilde G$ satisfies 
\begin{equation} \label{coloringcond}
\forall s \in \tilde V,\quad \bigcup_{v \in N(s) }\{ c(v)  \} = [1, \ell],
\end{equation}
the coloring is called a {\em valid $\ell$-coloring}.

It is evident that a color put on a node represents a message symbol associated to the node via 
a decoding function; namely, $\psi(v) = c(v)$. If we have a valid coloring, it is trivial to construct an 
encoder function $\phi$ satisfying the consistency condition $\psi(\phi(s,m))=m$. On the other hand,
a valid triple $(\tilde G, \phi, \psi)$ naturally leads to a valid coloring.

For given $\tilde G \subseteq G$, let
\begin{equation}
\gamma(\tilde G) \defeq \max \{\ell \mid  \exists \mbox{ a valid coloring on } \tilde G  \mbox{ with $\ell$-colors}\}.
\end{equation}
From the definition of $\gamma$, it should be noted that the writing capacity can be 
recast as 
\begin{equation} \label{ourprob}
C(G) = \max_{\emptyset \ne \tilde G \subseteq G} \gamma(\tilde G).
\end{equation}
Finding the largest $\ell$ satisfying the condition 
(\ref {coloringcond}) for a fixed $\tilde G$ is equivalent 
to computing the {\em domatic number} of $\tilde G$: The domatic number
of a graph $G$ is defined as the maximum number of node-disjoint dominating 
sets in $G$, and it is obvious that each color class in a valid 
$\ell$-coloring constitues a dominating set of $\tilde G$. Since it 
is allowed to choose an appropriate subgraph $\tilde G \subseteq G$ in 
our problem, we call the problem (\ref{ourprob}) 
the {\em SubDP problem}, which requires to identify 
the subgraph $\tilde{G}$ whose domatic number is the 
maximum. We also extend the terminology of domatic number to
its subgraph version. The {\em subgraph domatic number} is defined as 
the maximum of domatic numbers over all sugraphs.

\begin{figure}[htbp]
\begin{center}
\includegraphics[scale=0.35]{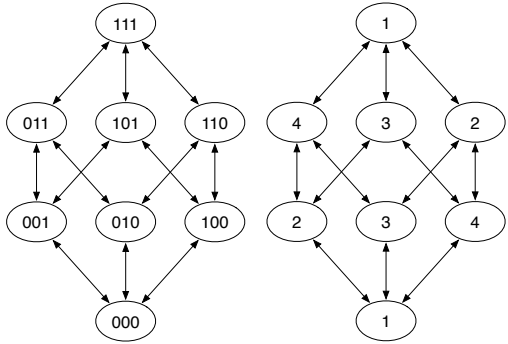}
\end{center}
\caption{A state transition graph and an optimal valid coloring: 
A node of the left graph corresponds to a vertex of 3-dimensional hypercube.
A binary 3-tuple in a node represents its coordinate. A pair of nodes corresponding to 
a pair of adjacent vertices has bidirectional edge connections.
This means that a state transition changes the binary state label only by one-bit.
(Right) Optimal valid coloring for this graph. We can prove that $C(G) = 4$ in this case.}
\label{validpartition1}
\end{figure}

\begin{example}
Let us consider a directed graph $G=(V,E)$ illustrated in Fig.\ref{validpartition1} (Left)
that represents the adjacency relationships of vertices of 3-dimensional hypercube.
Figure \ref{validpartition1} (Right) provides an optimal valid coloring $(G = \tilde G)$.
In this case, we can let ${\cal M} = [1,4]$. A state trajectory of an encoding process
can be seen as a walk on vertices along with edges of the 
hypercube.
\end{example}

\begin{figure}[htbp]
\begin{center}
\includegraphics[scale=0.35]{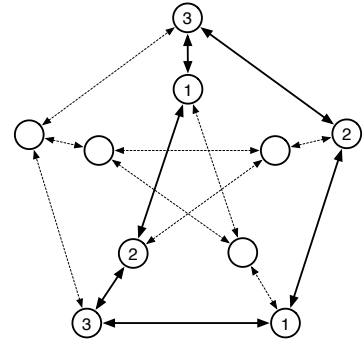}
\end{center}
\caption{An optimal subgraph of Petersen graph is denoted by solid lines.
The optimal valid coloring with 3-colors is indicated with the numbers form 1 to 3.}
\label{validpartition2}
\end{figure}

\begin{example}
Figure \ref{validpartition2} shows a non-trivial case. Petersen graph is a 3-regular graph 
with 10-nodes. In this case, we can prove that $C(G) = 3$.
\end{example}

\subsection{subDP is NP-complete} \label{completeness}
It is not difficult to show that the subgraph domatic partition 
problem is computationally hard. 

\begin{theorem}
The subDP problem is NP-complete even if the
input graph $G$ is $d$-regular for any $d \geq 3$. 
\end{theorem}
(Proof)
It is trivial that the problem belongs to NP. Thus what we have
to prove is the NP-hardness of the problem. The proof is the 
reduction from {\em full domatic partition problem for regular 
graphs}, which is the problem of deciding whether a given regular 
graph is {\em domatically full}, that is, the domatic number of a 
given $d$-regular graph is $d + 1$, or not. It is known 
that the full domatic partition problem is NP-complete for any 
$d \geq 3$~\cite{Kratochvil}. The core of the reduction is
the that the subgraph domitic number of any $d$-regular 
graph $G$ becomes $d+1$ if and only if for any $G$ is domatically 
full, which is verified by the following simple observation:
Since any subgraph $\tilde G \subset G$ contains a node with degree 
$d - 1$ or less, its domatic number cannot be greater than $d$.
Thus the sugraph domatic number of $G$ can become $d+1$ if and only
if $\tilde G = G$ and the domatic number of $\tilde G$ is $d + 1$,
which implies that $G$ is domatically full. \hfill\qed

It is worth pointing out that the computational complexity of the 
(standard) domatic number belongs to much harder class than NP-completeness. 
The domatic number is not only hard to compute the exact value, but also
hard to be approximated even within a factor better than $\ln n$~\cite{Feige}.
Since the subgraph domatic number appears to include more complex 
combinatorial choices, we conjecture that computing the subgraph domatic 
number also belongs to the class extremely harder than NP-completeness.
However, we unfortunately failed to prove the conjecture, and it is still 
open. Note that applying the strategy used in the proof for the 
(standard) domatic number is quite far from trivial. 

\section{Main contributions}

The following two theorems express  asymptotic behaviors of the
writing capacity $C(G)$ for bidirectional graphs. These results give us 
insight on the behavior of $C(G)$ for dense graph sequences.

\begin{theorem}[Dense graph sequence]
\label{dencetheorem}
Let  
\[
G_n = (V_n, E_n), n=1,2, \ldots
\]
be a sequence of bidirectional graphs
satisfying the following conditions.
The number of nodes $|V_n|$  is  assumed to be $n$ and 
the number of edges grows as 
$
|E| = \alpha n^2
$
where
$\alpha (0 < \gamma \le 1)$ is a real constant.
For such a sequence $G_n$, 
\begin{equation}
C(G_n) = \Omega\left( \frac{n}{\ln n} \right)
\end{equation}
holds in the regime where $n \rightarrow \infty$.
\end{theorem}

\begin{theorem}[Extremely dense graph sequence]
\label{extdencetheorem}
Let  
$
G_n = (V_n, E_n), \ n=1,2,\ldots
$
be a sequence of bidirectional graphs.
If the number of edges grows as 
\begin{equation}
|E_n| = {n^2} \left[ 1 - O\left(\frac 1 n \right) \right],
\end{equation}
then 
\[
C(G_n) = \Theta(n)
\]
holds in the regime where $n \rightarrow \infty$.
\end{theorem}

In the following sections, we will establish required lemmas 
to prove these theorems.

\section{Bounds on writing capacity}

In this section,  upper and lower bounds on $C(G)$ will 
be presented.

\subsection{Upper bound based on maximum degree}

The following upper bound on the writing capacity is simple but 
is fairly useful to determine the writing capacity of several special graph classes.
\begin{lemma}[Degree bound]
Suppose that a directed graph  $G=(V,E)$ is given.
The writing capacity $C(G)$ satisfies 
\begin{equation} \label{degbound}
C(G) \le 1 +\Delta_{out}(G).
\end{equation}
\end{lemma}
(Proof) 
Since $\tilde G =(\tilde V, \tilde E)$ is a non-empty subgraph of $G$, 
it is evident that 
$
|N(v) | = 1 +  d_{out}^{\tilde G}(v)
$
holds for any $v \in \tilde V$ and that $d_{out}^{\tilde G}(v) \le \Delta_{out}(G)$ for any 
$v \in \tilde V$.
This implies that the number of colors of a valid coloring
for any $\tilde V$ can be bounded by $1 +\Delta_{out}(G)$ from above. 
This fact directly leads to the claim of the lemma. \hfill\qed

This lemma clarifies the behavior of $C(G)$ for a sequence of 
sparse graphs. Let $G_n = (V_n,E_n)$ be a sequence of 
a sparse graphs that satisfies $\Delta_{out}({G_n}) = O(1)$.
The degree bound (\ref{degbound}) gives inequality 
$
C(G_n) \le \Delta_{out}({G_n}) + 1
$
that implies $C(G) = O(1)$ as well.

\subsection{Lower bound based on Lov\'asz local lemma}

In this subsection we will show the following lower bound on 
the writing capacity $C(G)$. The proof of this lemma is based on 
Lov\'asz local lemma that is a common technique to show 
the existence of valid configuration that satisfies all the requirements.
The outline of the following argument follows the proof of Lemma 2 of Feige et al. \cite{Feige}.
\begin{lemma}[LLL lower bound]
\label{LLLlemma}
Let $G=(V,E)$ be a directed graph and $\tilde G$ be any subset of $G$.
The writing capacity is bounded from below as
\begin{equation}\label{LLLlowerbound}
C(G) \ge \left\lfloor \frac{ \delta_{out}(\tilde G) +1}{ 1 +  \ln 3 + \ln 2 + 3 \ln\Delta } \right\rfloor,
\end{equation}
where $\Delta \defeq \Delta_{out}(G) + \Delta_{in}(G)$.
\end{lemma}
(Proof) 
We will put $\ell$-colors $(\ell \in [1, \Delta_{out}(G)+1 ])$ to nodes of $\tilde G$ uniformly randomly.
A {\em bad event} on $v \in \tilde G$ is the event such that there is a missing color 
in the neighborhood set $N(v)$. Precisely speaking, 
$
\{c(v) | v \in \tilde V\} \ne [1, \ell]
$
holds for a bad event on $v$, where $c(v) \in [1, \ell]$ represents the color put on the node $v$.
Assume that the probability $p(v)$ is the the probability of the bad event.
By using the union bound, the probability $p(v)$ can be upper bounded as follows:
\begin{eqnarray} \nonumber
p(v)
&\le& \sum_{i=1}^\ell Prob[\mbox{node with color } i \notin N(v)]  \\ 
&=& \ell \left(1 - \frac 1 \ell \right)^{|N(v)|}
\le \ell \exp \left(- \frac{|N(v)|}{\ell} \right).
\end{eqnarray}
We used the inequality $1-x \le \exp(-x)$ in the last step in the above derivation.
From this upper bound on $p(v)$, it is easy to derive a uniform upper bound.
Since $|N(v)| \ge \delta_{out}(\tilde G) + 1$,  we have 
$
p(v) \le p
$
for any $v  \in \tilde G$ where 
\begin{equation}
p \defeq (\Delta_{out}(G)+1) \exp \left(- \frac{ \delta_{out}(\tilde G)  +1}{\ell} \right).
\end{equation}

In order to use LLL, 
we need to have a upper bound on the number of bad events that are not 
mutually independent. 
Let us denote the number of such events by $T(v)$.
The number $T(v)$ can be bounded from above in the following way:
\begin{eqnarray} \nonumber
T(v) &\le& d_{out}(v) +d_{in}(v)   \\ \nonumber
&+& (d_{out}(v) + d_{in}(v)) (\Delta_{out}(\tilde G)+\Delta_{in}(\tilde G))  \\ \nonumber
&\le& 2\Delta^{2}  \defeq D.
\end{eqnarray}
This is because bad events on $v$ and $u$ are mutually independent
if the length of the shortest path between 
the nodes $v$ and $u$ in $\tilde G$ is greater than 2.

LLL guarantees the existence of a valid coloring if
\begin{equation} \label{LLL}
e p (D + 1) \le 1
\end{equation}
holds. The lefthand side $e p(D+1)$ can be evaluated as
\begin{eqnarray} \nonumber
e p (D+1) &=& e  (\Delta_{out}(G)+1) \exp \left(- \frac{\delta_{out}(\tilde G)  +1}{\ell} \right) (D + 1) \\ \nonumber
&\le&   \exp \left(- \frac{\delta_{out}(\tilde G)  +1}{\ell} + 1 + \ln(6 \Delta^3) \right).
\end{eqnarray}

We will consider the condition that the exponent becomes non-positive.
From the non-positiveness condition
\begin{equation}
- \frac{\delta_{out}(\tilde G)  +1}{\ell} + 1 + \ln(6 \Delta^3) \le 0,
\end{equation}
we have a sufficient condition:
\begin{equation}
\ell \le \frac{\delta_{out}(\tilde G)+1}{ 1 + \ln(6 \Delta^3) }.
\end{equation}
for existence of a valid coloring.
Namely, if $\ell$ is bounded as in the above inequality, 
the LLL condition (\ref{LLL}) is hold.
Since the number of colors $\ell$ is integer,  it is reasonable to let
\begin{equation} \label{LLLcond}
\ell = \left\lfloor \frac{\delta_{out}(\tilde G)+1}{ 1 + \ln(6 \Delta^3) } \right\rfloor
\end{equation}
because our aim is to maximize $\ell$. 
From LLL,  it is evident that the condition (\ref{LLLcond}) is a sufficient condition 
of existence of valid $\ell$-coloring for $\tilde G$.
\hfill\qed

\subsection{Proof of Theorem \ref{dencetheorem}}

The proof of Theorem \ref{dencetheorem} can be divided into two parts; 
in the first part of the proof, we will show that there is an
induced subgraph of a bidirectional graph $G$ that has minimum outbound degree larger than 
the half of average outbound degree of $G$.
Since the minimum outbound degree
restricts the number of colors in a valid coloring, finding an appropriate subgraph
of $G$ is crucial to increase $C(G)$.
In the second part of the proof, random coloring is employed and 
we shall use Lemma \ref{LLLlemma} to derive a sufficient condition in asymptotic regime.

We firstly will show that any bidirectional graph
has an induced subgraph with minimum outbound degree larger than 
half of the average outbound degree. 
A process to obtain an induced subgraph with large minimum outbound degree
is described in the following definition.
\begin{definition}[Subgraph sequence]
Let $G=(V,E)$ be a bidirectional graph.
A sequence of induced subgraphs of $G$, 
\[
G_0=G \supseteq G_1\supseteq G_2 \supseteq \cdots \supseteq G_K
\]
is recursively constructed in the following way.
If there exists a node $v \in G_i$ satisfying 
\begin{equation} \label{elim}
\frac 1 2 \epsilon_{out}(G_i) \ge d_{out}^{G_i}(v),
\end{equation}
then $G_{i+1}$ is obtained by removing both $v$ and the adjacent edges
from $G_i$.
The terminal subgraph $G_K$ does not contain a node satisfying  
the condition (\ref{elim}).
The sequence constructed in such a way is said to be appropriate.
\end{definition}

The next lemma states that the terminal subgraph $G_K$ is what we want.
\begin{lemma}\label{induced}
Let $G=(V,E)$ be a balanced directed graph. 
Suppose also that an appropriate sequence of induced subgraphs of $G$,
$G_0 \supseteq G_1\supseteq G_2 \supseteq \cdots \supseteq G_K$, is given.
The minimum outbound degree of $G_K$ satisfies the inequality 
$
\delta_{out}(G_K) >  {\epsilon_{out}(G) }/{2}.
$
\end{lemma}
The proof of this lemma is given in \cite{Dielstel}.

We are ready to go into the proof of Theorem \ref{dencetheorem}.
Let $G = (V,E)$ be a bidirectional graph. 
A subgraph $\tilde G$ can be chosen arbitrary so that $\delta_{out}(\tilde G)$ 
becomes large. We here take the following strategy to choose $\tilde G$:
If $\delta_{out}(G)> \epsilon_{out}(G)/2$, then let $\tilde G= G$.
Otherwise, namely $\delta_{out}(G) \le \epsilon_{out}(G)/2$, 
then let $\tilde G= G_K$ where $G_K$ is
the terminal subgraph in an appropriate sequence of induced subgraphs 
$G_0 \supseteq G_1\supseteq G_2 \supseteq \cdots \supseteq G_K$.
According to this choice, we obtain
\begin{equation} \label{outdegree}
\delta_{out}(\tilde G) \ge \max \{\delta_{out}(G), \epsilon_{out}(G)/2  \}
\end{equation}
due to Lemma \ref{induced}.

Combining the condition (\ref{LLLcond}) and (\ref{outdegree}), we immediately have 
a sufficient condition for $\ell$-valid coloring:
\begin{equation}
\ell = \left\lfloor \frac{ \max \{\delta_{out}(G), \epsilon_{out}(G)/2  \} +1}
{ 1 + \ln 3 + \ln 2 + 3 \ln  (\Delta_{out}(G) ) } \right\rfloor.
\end{equation}
Note that $\Delta = 2 \Delta_{out}(G)$ holds for a bidirectional graph.

Assume that $G_n=(V_n,E_n)$ is a sequence of bidirectional graph
with $|E_n| = \alpha n^2$.
The lefthand side of (\ref{LLLlowerbound}) can be lower bounded  as
\begin{equation}
\left\lfloor \frac{ \max\{\epsilon_{out}(G_n)/2, \delta_{out}(G_n) \} +1}{A+ 3 \ln(\Delta_{out}) } \right\rfloor
\ge \frac{\alpha n/2+ 1}{ A+ 3 \ln n } -1,
\end{equation}
where
$
A \defeq  1 +  \ln 3 + \ln 2.
$
It is clear that we have
$
C(G) = \Omega\left({n}/{\ln n} \right)
$
under the condition $n \rightarrow \infty$.
This completes the proof of Theorem \ref{dencetheorem}.

As a byproduct of the LLL lower bound, 
a polynomial time approximation algorithm for evaluating the writing capacity 
with approximation ratio $O(\ln \Delta_{out}(G) )$ can be derived where $\Delta_{out}(G)$ is the maximum 
outbound degree of $G$.

\subsection{Lower bounds on writing capacity based on Tur\'an's lemma}

A graph with extremely high edge density contains a subgraph that 
have a large writing capacity such as complete graphs and bipartite graphs.
If a given graph $G$ includes such a subgraph, its writing capacity can be 
lower bounded by the writing capacity of the subgraph.
The result shown in this subsection is based on the Tur\'an's lemma 
that is a fundamental result in extremal graph theory.

We firstly start our argument on this subsection with the following basic lower bound.
\begin{lemma}[Subgraph lower bound] \label{monotone}
Let $G=(V,E)$ be a directed graph and $\tilde G$ be a subgraph of $G$.
We have
$
C(G) \ge C(\tilde G).
$

\end{lemma}
(Proof) The subgraph $\tilde G$ includes a subgraph $G' (G' \subseteq \tilde G \subseteq G)$
with $C(\tilde G)$-color valid coloring. Because the subgraph $G'$ is also a subgraph of the graph $G$,
it is clear that $C(\tilde G)$ becomes a lower bound of $C(G)$.
\hfill\qed

Tur\'an graph is the undirected graph such that 
it has the largest number of edges among undirected graphs with $n$-vertices which 
does not contain a complete graph $K_n$ as a subgraph.
The number of edges of Tur\'an graph is denoted by $t_{r-1}(n)$.
Tur\'an proved the following upper bound on the size of the Tur\'an graph.
\begin{lemma}[Tur\'an's lemma]
\begin{equation}
t_{r-1}(n) \le \frac 1 2 n^2 \frac{r-2}{r-1}.
\end{equation}
\end{lemma}

The upper bound of the edge size of Tur\'an graph leads to the 
following lemma that guarantees the existence of a large clique for an extremely dense graph. 
This lemma constitutes a main part of the proof of Theorem  \ref{extdencetheorem}.

\begin{lemma}\label{turantheorem}
Let $G=(V,E)$ be a bidirectional graph.
If 
\begin{equation}\label{turan}
|E| > \frac 1 2 n^2 \frac{\lfloor \alpha n  \rfloor-2}{\lfloor \alpha n \rfloor -1}
\end{equation}
holds, then $C(G)$ is lower bounded as
$
C(G) \ge \lfloor \alpha n \rfloor
$
where $\alpha(0 < \alpha \le 1)$ is a real constant.
\end{lemma}
The proofs of this lemma and Theorem \ref{extdencetheorem} are omitted.
The forthcoming extended version of this paper will include these proofs and also discussion on 
a polynomial time approximation algorithm for subDP problems.

We will firstly show that the complete graph $K_n$ has the writing capacity 
$
C(K_n) = n.
$
For each node of $K_n$, we will put a distinct color from $[1,n]$.
It is clear that this coloring is a valid coloring for $K_n$. Thus, we have $C(K_n) \ge n$.
On the other hand, due to the degree bound, we obtain another inequality $C(K_n) \le n$.
These two inequalities imply $C(K_n) = n$.

From Tur\'an's lemma, if the condition (\ref{turan}) is met, then the graph $G$
contains $K_{\lfloor \alpha n \rfloor}$ as a subgraph. By Lemma \ref{monotone}, 
we finally have $C(G) \ge \lfloor \alpha n \rfloor$. \hfill\qed

The proof of Theorem \ref{extdencetheorem} is given as follows.
The coefficient appeared in (\ref{turan}) can be bounded from above:
\begin{equation}
\frac{\lfloor \alpha n  \rfloor-2}{\lfloor \alpha n \rfloor -1} = 1 - \frac{1}{\lfloor \alpha n \rfloor - 1} < 1 - \frac{1}{\alpha n}.
\end{equation}
Due to the assumption on the number of edges,
there exists a positive number $C$ satisfying 
$
|E| \ge {n^2}/{2} \left(1 -  {C}/{n}  \right)
$
for sufficiently large $n$. By letting $\alpha \defeq 1/C$, 
Lemma \ref{turantheorem} guarantees 
$
C(G) \ge \alpha n.
$
On the other hand, the degree bound givens $C(G) = O(n)$.
Combining these, we immediately obtain $C(G) = \Theta(n)$  in
the asymptotic regime.

\section{Approximation algorithm}
As proved in Subsection \label{completeness},  SubDP problems are NP-complete.
It is natural to try developing an approximation algorithm 
to solve a subDP problem that is equivalent to the evaluation of the writing capacity of a given graph.
In section, we will show that evaluation of the writing capacity of a bidirectional graph 
with the approximation ratio $O(\ln \Delta_{out}(G))$ can be done in polynomial time.
The key to derive the approximation ratio is to use a tight upper bound that can be 
computed in polynomial time with $n$.
Let $G$ be a bidireced graph with $n$ nodes. 
A $k$-{\em core} of $G$ is a connected subgraph $G$ with 
the minimum outbound degree $k$. It is known that $k$-core can be found in linear time
if it exists.
The following upper bound is based on evaluation of $k$-cores of $G$.
\begin{lemma}[$k$-core bound]
Let $G$ be a bidirectional graph and $\eta(G)$ be
$
\eta(G) \defeq \max \{k | G \mbox{ has a $k$-core}  \}.
$
The writing capacity $C(G)$ is upper bounded as 
\begin{equation} \label{kcorebound}
C(G) \le 1+ \eta(G).
\end{equation}
\end{lemma}
(Proof) 
The writing capacity $C(G)$ can be bounded from above as
\begin{eqnarray}
C(G) 
&=& \max_{\emptyset \ne \tilde G \subseteq G} \gamma(\tilde G) \\
&\le& \max_{\emptyset \ne \tilde G \subseteq G} [1 + \delta_{out}(\tilde G)] 
\le 1 + \eta(G).
\end{eqnarray}
The last inequality is due to the definition of $\eta(G)$. \hfill\qed

The process of our approximation algorithm is summarized as follows.
At the first stage of the algorithm, we will find a subgraph $\tilde G$ that
satisfies $\delta_{out}(\tilde G)=\eta(G)$.
Such a subgraph $\tilde G$ can be obtained in polynomial time.
In the second stage of the algorithm, an algorithmic 
version of LLL with setting
\begin{equation}\label{lowe}
 \ell^* \defeq \left\lfloor \frac{ 1+\eta(G)}{ 1 +  \ln 3 + \ln 2 +  3 \ln\Delta } \right\rfloor
\end{equation}
is exploited. It can find a valid coloring with $\ell^*$-colors in polynomial time.
Closely related approximation algorithms for domatic partition problem is intensively 
discussed in \cite{Feige}.
Combining the upper bound (\ref{kcorebound}) and the lower bound (\ref{lowe}),
we immediately have the following upper bound on the approximation ratio
$
{C(G)}/{\ell^*} = O(\ln \Delta_{out}(G)),
$
which is achieved with the above algorithm in polynomial time.

\section*{Acknowledgement}
The first author would like to express his sincere appreciation to
Hiroshi Kamabe and Kees Immink for their constructive comments for 
an earlier version of this work.
This work is supported 
by JSPS Grant-in-Aid for Scientific Research (B) Grant Number 25289114.

\end{document}